\documentclass[12pt,a4]{article}
\usepackage{amsmath}
\usepackage{bm}
\usepackage{amsfonts}
\usepackage{amssymb}
\usepackage{graphics}
\usepackage[normal]{caption2}
\usepackage{subfigure}
\usepackage{rotating}
\usepackage{citesort}
\def\be{\begin{equation}}
\def\ee{\end{equation}}
\def\ba{\begin{array}}
\def\ea{\end{array}}
\def\bea{\begin{eqnarray}}
\def\eea{\end{eqnarray}}

\begin{document}
\baselineskip 20pt \setlength\tabcolsep{2.5mm}
\renewcommand\arraystretch{1.5}
\setlength{\abovecaptionskip}{0.1cm}
\setlength{\belowcaptionskip}{0.5cm}
\pagestyle{empty}
\newpage
\pagestyle{plain} \setcounter{page}{1} \setcounter{lofdepth}{2}
\begin{center} {\large\bf  Role of isospin degree of freedom on N/Z dependence of participant-spectator matter}\\
\vspace*{0.4cm}

{\bf Sakshi Gautam} \footnote{Email:~sakshigautm@gmail.com}\\
{\it  Department of Physics, Panjab University, Chandigarh -160
014, India.\\}
\end{center}

\section*{Introduction}
With the availability of high-intensity radioactive beams at many
facilities as well as a number of next generation beam facilities
being constructed or being planned, the studies on the role of
isospin degree of freedom have recently attracted a lot of
attention in both nuclear physics and astrophysics. The ultimate
goal of such studies is to extract information on the isospin
dependence of in-medium nuclear effective interactions as well as
equation of state (EOS) of isospin asymmetric nuclear matter. The
later quantity especially the symmetry energy term is important
not only to nuclear physics community as it sheds light on the
structure of radioactive nuclei, reaction dynamics induced by rare
isotopes but also to astrophysics community as it acts as a probe
for understanding the evolution of massive stars and supernova
explosion. Role of isospin degree of freedom has been investigated
in collective flow and its disappearance at balance energy
\cite{pak97,gaum1}. At balance energy, attractive interactions due
to mean field are balanced by repulsive interactions due to
nucleon-nucleon collisions and this counterbalancing is reflected
in quantities like participant-spectator matter \cite{sood}. In
the present work, we study the role of isospin degree of freedom
on the N/Z dependence of participant-spectator matter. The study
is carried out within the framework on isospin-dependent quantum
molecular dynamics model \cite{hart98} which is the extension of
quantum molecular dynamics model.  The IQMD model
 treats different charge states of
nucleons, deltas, and pions explicitly, as inherited from the
Vlasov-Uehling-Uhlenbeck (VUU) model.

\begin{figure}[!t] \centering
\vskip 0.5cm
\includegraphics[angle=0,width=12cm]{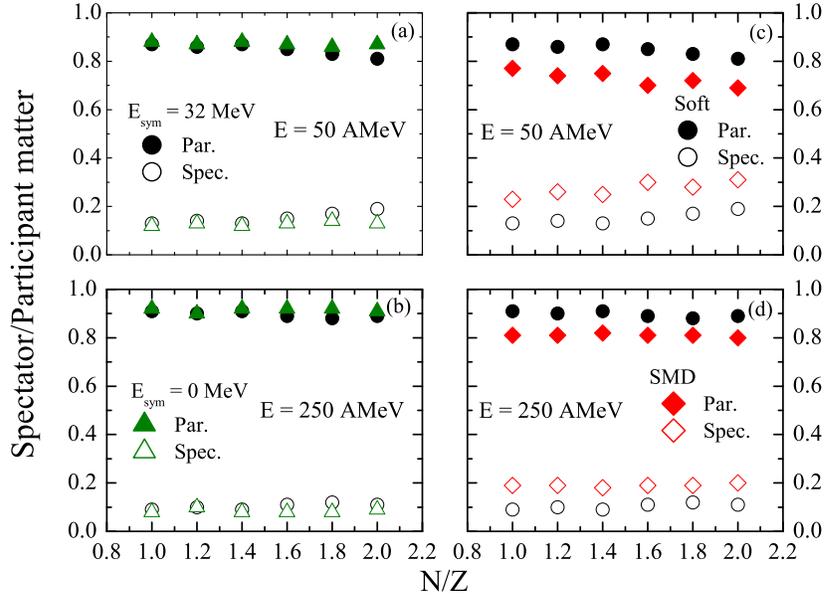}
\vskip -0.5cm \caption{ The N/Z dependence of participant and
spectator matter for the reactions of Ca+Ca at 50 (upper panels)
and 250 (lower) MeV/nucleon for different N/Z ratios with
E$_{sym}$ = 0 MeV (left panels) and MDI (right). Various symbols
are explained in the text.}\label{fig3}
\end{figure}

\section*{Results and discussion}
We simulate the reactions of Ca+Ca series series having
 N/Z ratio varying from 1.0 to 2.0 in steps of 0.2. In particular we simulate the
 reactions of $^{40}$Ca+$^{40}$Ca, $^{44}$Ca+$^{44}$Ca,
 $^{48}$Ca+$^{48}$Ca, $^{52}$Ca+$^{52}$Ca, $^{56}$Ca+$^{56}$Ca,
 and $^{60}$Ca+$^{60}$Ca at b/b$_{max}$ = 0.2-0.4. The incident
 energies are taken to be 50 and 250 MeV/nucleon. The reactions are followed till the transverse flow
 saturates. Since the E$_{bal}$ is smaller in heavier colliding nuclei compared to the
 lighter ones, so the saturation of transverse flow occurs much earlier
 in lighter colliding systems as compared to the heavier ones. Though
 transverse flow saturates much earlier, there are certain variables
 which keep on changing, so we follow all the reactions uniformly upto 800 fm/c.
\par
In fig. 1, we display the N/Z dependence of participant-spectator
matter for Ca+Ca reactions (circles). Solid (open) symbols are for
participant (spectator) matter. Upper (lower) panels represent the
results for 50 (250) MeV/nucleon. From figure, we find that at
participant-spectator matter changes slightly with N/Z ratio at
lower incident energy of 50 MeV/nucleon. On the other hand, the
participant-spectator matter is almost independent of the N/Z
ratio at 250 MeV/nucleon. When we increase the N/Z ratio at fixed
Z, the mass of the system increases. Due to increase in mass, the
participant matter should increase with N/Z (at fixed incident
energy). Also, as we moving to higher N/Z ratios, the role of
symmetry energy also increases. And due to repulsive nature of
symmetry energy, lesser number of collisions will take place at
central dense zone and hence amount of participant matter should
decrease with increase in N/Z ratio. Hence, the net effect is due
to the interplay of mass effects and symmetry energy effects. To
see the relative importance of these two effects, as a next step,
we make the strength of symmetry energy zero and calculate the
participant-spectator matter for Ca+Ca reactions. The results are
displayed in Fig. 1(a) (50 MeV/nucleon) and 1(b) (250 MeV/nucleon)
(triangles). From figure, we see that on reducing the strength of
symmetry energy to zero, the participant (spectator) matter
increases (decreases) slightly for higher N/Z ratios whereas the
effect on lower N/Z ratios is almost negligible. This indicates
that the decrease in participant matter for higher N/Z ratios is
due to the role of symmetry energy. The momentum-dependent
interactions play significant role in the dynamics of heavy-ion
collisions. To see the role of MDI on participant-spectator matter
we calculate the participant-spectator matter for SMD EOS for the
reactions of Ca+Ca. The results are displayed by diamonds in Fig.
1(c) and 1(d) at 50 and 250 MeV/nucleon, respectively. From
figure, we see that participant [closed diamonds] (spectator, open
diamonds) matter decreases (increases) with SMD. This is due to
the fact that since MDI is repulsive in nature, so the matter is
thrown away from the central dense zone and hence lesser number of
collisions will take place which leads to decrease (increase) in
participant (spectator) matter.

\section*{Acknowledgments}
 This work has been supported by a grant from Centre of Scientific
and Industrial Research (CSIR), Govt. of India.


\begin{thebibliography}{999}

\bibitem{pak97} R. Pak \emph{et al}., Phys. Rev. Lett. \textbf{78}, 1022 (1997);\emph{ ibid}.
\textbf{78}, 1026 (1997).

\bibitem{gaum1} S. Gautam \emph{et al}., J. Phys G: Nucl. Part.
Phys. \textbf{37}, 085102 (2010).
\bibitem{sood} A. D. Sood\emph{ et al}., Phys. Rev. C \textbf{70}, 034611 (2004);
\emph{ibid.} C\textbf{79}, 064618 (2009).

\bibitem{hart98} C. Hartnack \emph{et al}., Eur. Phys. J. A
\textbf{1}, 151 (1998); R. K. Puri \emph{et al}., Nucl. Phys. A
\textbf{575}, 733 (1994); E. Lehmann \emph{et al}., Phys. Rev. C
\textbf{51}, 2113 (1995); R. K. Puri \emph{et al}., J. Comp. Phys.
\textbf{162}, 245 (2000).


\end{thebibliography}
\end{document}